\documentstyle[12pt,fleqn]{article}

\catcode`\@=11
\@addtoreset{equation}{section}

\begin{document}
\begin{flushright}
OU-HET/206\\hep-th/9411???\\November 1994
\end{flushright}
\vspace{0.5in}
\begin{center}\Large{\bf \lq\lq Moduli Space" of Asymptotically
Anti-de Sitter Spacetimes in (2+1)-Dimensions} \\
\vspace{0.2in}
\vspace{1cm}\renewcommand{\thefootnote}{\fnsymbol{footnote}}
\normalsize\ Kiyoshi Ezawa\footnote[1]{Supported by JSPS.}
\footnote[2]{e-mail address:
ezawa@funpth.phys.sci.osaka-u.ac.jp}\setcounter{footnote}{0}

\vspace{0.5in}

        Department of Physics \\
        Osaka University, Toyonaka, Osaka 560, Japan\\
\vspace{0.1in}
\end{center}
\vspace{0.5in}
\baselineskip 17pt
\begin{abstract}
Setting an ansats that the metric is expressible by
a power series of the inverse radius and taking a particular
gauge choice, we construct a \lq\lq general solution"
of (2+1)-dimensional Einstein's equations with a
negative cosmological constant in the case
where the spacetime is asymptotically anti-de Sitter.
Our general solution turns out to be parametrized by
two centrally extended quadratic
differentials on $S^{1}$. In order to include 3-dimensional
Black Holes naturally into our general solution,
it is necessary to exclude the region inside the horizon.
We also discuss the relation of our general solution to the moduli
space of
flat $\tilde{SL}(2,{\bf R})\times\tilde{SL}(2,{\bf R})$ connections.\\
                    \\
classification numbers (PACS nos.): 04.20, 04.60
\end{abstract}
\newpage

\baselineskip 20pt

\section{Introduction}
 \ \ \ \
When we try to quantize General Relativity (GR),
we usually run into serious obstructions, which include
the \lq\lq issue of time in quantum gravity" and the
\lq\lq problem of finding (local) observables"\cite{dewitt}.
These two obstructions are both closely related to the
general covariance of GR.

To see from the viewpoint of the canonical formalism,
in the case of pure gravity on a compact spatial manifold,
 the Hamiltonian of GR
is expressed by a linear combination of the first class constraints.
The time evolution of the canonical variables via the
Hamiltonian is nothing but a gauge transformation which cannot
be observed physically. These first class constraints
generates general coordinate transformations of the spacetime,
which prevents us from making a distinction between

a point on spacetime and another point.

One of the ways to circumvent these difficulties is to consider
spacetimes which are
asymptotically isometric to some well-behaved space ( such as
Minkowski, de Sitter, or anti-de Sitter space). The Hamiltonian
defined on these space has a nontrivial contribution from
the boundary, according to which we have a possibility to define
a meaningful time evolution \cite{RT}. Moreover, we can construct
physical observables at spatial infinity because the asymptotic
condition restricts the types of diffeomorphisms
allowed at the infinity. It would therefore be
important to investigate GR on a spacetime which is
asymptotically flat (or (anti-) de Sitter).

In this paper we investigate the asymptotically anti-de Sitter
spacetimes in (2+1)-dimensions. In (2+1)-dimensions, Einstein's
equations
with a negative cosmological constant  tell us that the spacetimes
be locally anti-de Sitter ($ADS^{3}$)\cite{DJ}.

If we consider naively from this fact, solutions of Einstein's
equations
which are asymptotically $ADS^{3}$seem to be exhausted by
3-dimensional black holes (3DBH)\cite{BTZ} possibly with a negative
mass.
The main purpose of this
paper is to investigate whether this is indeed the case.

In \S 2 while reviewing the canonical formalism of asymptotically
$ADS^{3}$ spacetimes\cite{brown} in terms of Chern-Simons
formulation of GR\cite{town}\cite{witte},
we show that the diffeomorphism equivalence classes of
asymptotically $ADS^{3}$ spacetimes are characterized by two
centrally extended quadratic differentials on $S^{1}$.
In \S 3 we solve Einstein's equations explicitly by imposing a
particular
gauge-fixing condition and by making
an ansatz that the metric should be expressible by
a power series of inverse radial coordinate.
3DBH's turn out to be naturally involved in our general solution if
we neglect the region inside the outer horizon.
To obtain some intuition about our general solution, we investigate
some simple cases in \S 4. While we find new solutions which
do not belong to 3DBH, these solutions appear to
be physically irrelevant because they involve closed timelike curves.
\S 5 is devoted to the analysis of topological structure of
the moduli space. Its relation to the moduli space of flat
$\tilde{SL}(2,{\bf R})\times\tilde{SL}(2,{\bf R})$ connections is also
suggested.
In \S 6, after summarizing the main results, we discuss the remaining
issues
on the asymptotically $ADS^{3}$ spacetimes.


\section{Effective Theory of Asymptotically
Anti-de Sitter Spacetimes}

We work in the first-order Einstein gravity
in (2+1)-dimensions with a negative
cosmological constant $\Lambda=-1/l^{2}$, which is shown to be

equivalent to the ${\rm SO}(2,2)$ Chern-Simons gauge theory
\cite{town}\cite{witte}. We use as fundamental variables
the triad $e^{a}=e_{\mu}^{a}dx^{\mu}$ and the spin connection
$\omega^{ab}=\omega_{\mu}^{ab}dx^{\mu}$.
\footnote{Our convention for the indices and the signatures of the
metrics
is the following:
$\mu,\nu,\rho,\cdots(=t,r,\phi)$ denote 2+1 dimensional spacetime
indices
and the metric $g_{\mu\nu}$ has the signature $(-,+,+)$;
$i,j,k,\cdots$ are  used for spatial indices;
$a,b,c,\cdots$ represent indices of the
local Lorentz group, with the metric $\eta_{ab}={\rm diag}(-,+,+)$;
$\epsilon_{abc}$ is the totally antisymmetric pseudo-tensor with
$\epsilon_{012}=-\epsilon^{012}=1$.}
If we assume that the spacetime manifold $M$ has a boundary $\partial
M$,
the action is
\begin{eqnarray}
I &=& \int_{M}\epsilon_{abc}e^{a}\wedge
[d\omega^{bc}+\omega^{b}_{\mbox{ }d}\wedge\omega^{dc}-
\frac{1}{3}\Lambda e^{b}\wedge e^{c}]	+B^{\prime}(\partial M)
	\nonumber \\*
&=&\int_{M}{\rm Tr}[A\wedge dA+\frac{2}{3}A\wedge A\wedge A]
+B^{\prime}(\partial M),
\label{eq:EPac}
\end{eqnarray}
where $B^{\prime}(\partial M)$ is the boundary term which is necessary
for the variational principle to give local equations of motion.
We have also introduced SO$(2,2)$ connection $A\equiv P_{a}e^{a}
+\frac{1}{2}\epsilon_{abc}J^{a}\omega^{bc}$ with $(J_{a},P_{a} )$
being the generators of SO$(2,2)$ Lie algebra:
$$
[J_{a},J_{b}]=\epsilon_{abc}J^{c},
[J_{a},P_{b}]=\epsilon_{abc}P^{c},
[P_{a},P_{b}]=\frac{1}{l^{2}}\epsilon_{abc}J^{c}.
$$
Tr in eq.(\ref{eq:EPac}) denotes an invariant bilinear form on
SO$(2,2)$:
$$
{\rm Tr}(J_{a}P_{b})=\eta_{ab},\quad {\rm Tr}(J_{a}J_{b})=
{\rm Tr}(J_{a}J_{b})=0.
$$
The action in the canonical formalism is obtained by performing
the 2+1 decomposition $M\approx {\bf R}\times\Sigma$, $A=A_{t}dt+
\tilde{A}$ and $d=dt\partial_{t}+\tilde{d}$:
\footnote{$\partial_{t}$ denotes a partial differentiation with
respect to $t$. Tilde refers to
the quantities or the operations defined on $\Sigma$.}
\begin{eqnarray}
I &=&\int dt[\int_{\Sigma}{\rm
Tr}(-\tilde{A}\wedge\partial_{t}\tilde{A})-H],
\nonumber \\*
H&\equiv&-2\int_{\Sigma}{\rm
Tr}[A_{t}(\tilde{d}\tilde{A}+\tilde{A}\wedge
\tilde{A})]+B(\partial\Sigma).\label{eq:CSac}
\end{eqnarray}
We should notice that the boundary term $B(\partial\Sigma)$
of the Hamiltonian $H$ is introduced in order to make Hamilton's
principle
well-defined. This $B(\partial\Sigma)$ is determined by the following
functional differential equation
\begin{equation}
\int dt\delta B(\partial\Sigma)=-2\int_{\partial M}{\rm Tr}(A_{t}dt
\wedge\delta\tilde{A})=2\int dt\oint_{\partial\Sigma}d\phi
{\rm Tr}(A_{t}\delta A_{\phi}). \label{eq:hamiltonian}
\end{equation}

As is well known this system is a first class constraint system.
Assume we take the gauge-fixing method in which we explicitly solve the
constraints by imposing particular gauge-fixing conditions as many as
the number of the first class constraints.
The spatial part $\tilde{A}$ of the connection, which are the dynamical
degrees of freedom in the unconstrained system, are thus determined
by solving the constraint
$\tilde{F}\equiv\tilde{d}\tilde{A}+\tilde{A}\wedge
\tilde{A}=0$. While the temporal part $A_{t}$ remains as gauge degrees
of freedom, we can represent it in terms of $\tilde{A}$
by solving the equations of motion $F_{ti}=0$. The problem  thus
reduces to that of solving the equations of motion
\begin{equation}
F\equiv dA+A\wedge A=\frac{1}{2}F_{\mu\nu}
dx^{\mu}\wedge dx^{\nu}=0
\label{eq:EOM}
\end{equation}
under a particular gauge choice. We mention that this equations
of motion involve Einstein's equations $d\omega^{ab}+\omega^{a}
_{\mbox{  }c}
\wedge\omega^{cb}=0$ and torsion-free conditions
$de^{a}+\omega^{a}_{\mbox{  }b}\wedge e^{b}=0$.

Next we determine  the asymptotic form of $A$.
The condition that the metric should be asymptotically $ADS^{3}$ is
given by \cite{brown}:
\begin{eqnarray}
ds^{2}&=&-(\frac{r^{2}}{l^{2}}+O(1))dt^{2}+(\frac{l^{2}}{r^{2}}+
O(\frac{1}{r^{4}}))
dr^{2}+(r^{2}+O(1))d\phi^{2} \nonumber \\*
{}{}&&+O(1)dtd\phi+O(\frac{1}{r^{3}})drdt+O(\frac{1}{r^{3}})drd\phi.
\label{eq:aspmetric}
\end{eqnarray}
Due to the gauge degrees of freedom associated with
the local Lorentz transformations,
there are an infinitely many number of
SO$(2,2)$ connections $A$ which
give the above metric. By loosely fixing the local Lorentz gauge
degrees of
freedom, we put the following asymptotic condition on $A$:
\begin{equation}\begin{array}{l}
A=P_{0}l\{(\frac{r}{l}-\frac{Ml}{2r}+O(\frac{1}{r^{2}}))\frac{dt}{l}
+O(\frac{1}{r^{4}})dr+O(\frac{1}{r^{2}})d\phi\}\\
\quad+P_{1}l\{(\frac{1}{r}+\frac{M^{\prime}l^{2}}{2r^{3}}
+O(\frac{1}{r^{4}}))dr+O(\frac{1}{r^{2}})dt+O(\frac{1}{r^{2}})d\phi\}\\
\quad+P_{2}l\{(\frac{r}{l}+\frac{(M-M^{\prime})l}{2r}+O(\frac{1}{r^
{2}}))
d\phi+O(\frac{1}{r^{4}})dr+(-\frac{J}{2r}+O(\frac{1}{r^{2}}))
\frac{dt}{l}\}\\
\quad+J_{0}\{(\frac{r}{l}-\frac{Ml}{2r}+O(\frac{1}{r^{2}}))d\phi
+O(\frac{1}{r^{4}})dr+O(\frac{1}{r^{2}})dt\} \\
\quad+J_{1}\{(\frac{Jl}{2r^{3}}+O(\frac{1}{r^{4}}))dr
+O(\frac{1}{r^{2}})dt+O(\frac{1}{r^{2}})d\phi\}.\\
\quad+J_{2}\{(\frac{r}{l}+\frac{(M-M^{\prime})l}{2r}+O(\frac{1}{r^
{2}}))
\frac{dt}{l}+O(\frac{1}{r^{4}})dr+(-\frac{J}{2r}
+O(\frac{1}{r^{2}}))d\phi\}.\end{array}\label{eq:asp}
\end{equation}
In setting the above asymptotic form we have solved the equations
of motion(\ref{eq:EOM}) asymptotically. $M$, $M^{\prime}$ and $J$
in eq.(\ref{eq:asp}) correspond to the $O(1)$-part of $g_{tt}$,
$O(\frac{1}{r^{4}})$-part of $g_{rr}$ and $O(1)$-part of $g_{t\phi}$,
respectively.

According to ref.\cite{brown}, the group of the asymptotic symmetries
which preserve boundary condition (\ref{eq:aspmetric}) is isomorphic
to the pseudo-conformal group in 2 dimensions, which in turn is
isomorphic to the direct product group of two Virasoro groups.
In our formulation, these transformations are
generated by the SO$(2,2)$ gauge transformation $\delta_{\xi}A=d\xi
+[A,\xi]$. The gauge parameter $\xi$ is of the following form
\begin{equation}\begin{array}{lll}
\xi=P_{0}(B^{0}r+\frac{C^{0}}{r}+O(\frac{1}{r^{2}}))&+&
J_{0}(\frac{B^{2}}{l}r+\frac{D^{0}}{r}+O(\frac{1}{r^{2}}))\\
\quad+P_{1}(\beta^{1}+O(\frac{1}{r^{2}}))&+&
J_{1}(\gamma^{1}+O(\frac{1}{r^{2}}))\\
\quad+P_{2}(B^{2}r+\frac{C^{2}}{r}+O(\frac{1}{r^{2}}))&+&
J_{2}(\frac{B^{0}}{l}r+\frac{D^{2}}{r}+O(\frac{1}{r^{2}})),
\end{array}\label{eq:asgauge}
\end{equation}
with the coefficients subject to the relations:
\begin{equation}\begin{array}{lll}
\partial_{\phi}B^{2}=l\partial_{t}B^{0}=-\beta^{1}/l&,&
\partial_{\phi}B^{0}=l\partial_{t}B^{2}=-\gamma^{1}\\
C^{2}+lD^{0}=-\frac{M^{\prime}l^{2}}{2}B^{2}-\frac{Jl}{2}B^{0}&,&
C^{0}+lD^{2}=-\frac{M^{\prime}l^{2}}{2}B^{0}-\frac{Jl}{2}B^{2}\\
C^{2}-lD^{0}=l^{2}(\frac{2M-M^{\prime}}{2}-\partial_{\phi}^{2})B^{2}
-\frac{Jl}{2}B^{0}&,&
lD^{2}-C^{0}=l^{2}(\frac{2M-M^{\prime}}{2}-
\partial_{\phi}^{2})B^{0}-\frac{Jl}{2}B^{2}.
\end{array}\label{eq:g-coef}
\end{equation}
The $O(\frac{1}{r^{2}})$ terms depend on more detailed information on
the
connection $A$ which are not explicitly written in eq.(\ref{eq:asp}).

In the canonical formalism, the generator of the asymptotic gauge
transformation (\ref{eq:asgauge}) is expressed by a linear combination
of the first class constraints plus a surface term
\footnote{In practice these generators generates under the Poisson
bracket the transformation:
$$ \{A,G[\xi]\}_{P.B.}=\delta_{\xi}A+(\mbox{terms linear in the
constraints
with coefficients }\{A,\xi\}_{P.B.}).$$
The second term in the R.H.S., however, vanishes because we consider
the
constraints to be solved.}
\begin{eqnarray}
G[\xi]&=&-2\int_{\Sigma}{\rm Tr}[\xi(\tilde{d}\tilde{A}+\tilde{A}\wedge
\tilde{A})]+\int_{\Sigma}d^{2}x{\rm
Tr}[(\partial_{t}\xi+[A_{t},\xi])\Pi]
+Q[\xi],\nonumber \\*
\delta Q[\xi]&=&2\oint_{\partial\Sigma}{\rm Tr}(\xi \delta \tilde{A}),
\label{eq:generator}
\end{eqnarray}
where $\Pi$ denotes the conjugate momentum of $A_{t}$ and $\Pi\approx0$
also gives first class constraints.

{}From now on we will consider that the
boundary $\partial\Sigma$ of  the spatial manifold
consists only of the spatial infinity at $r\rightarrow\infty$.
The charge $Q[\xi]$ in this case  is given by
\begin{equation}
Q[\xi]=2\oint{\rm Tr}[\xi(\tilde{A}-(\tilde{A})_{0})],
\end{equation}
where $(A)_{0}$ is a fiducial connection. If we use as $(A)_{0}$ the
\lq\lq vacuum configuration" of 3DBH\cite{BTZ}
\begin{equation}
(A)_{0}=P_{0}\frac{r}{l}dt+P_{1}\frac{l}{r}dr+P_{2}rd\phi
+J_{0}\frac{r}{l}d\phi+J_{2}\frac{r}{l^{2}}dt,
\end{equation}
the charge is explicitly given by
\begin{equation}
Q[\xi] =\oint d\phi[B^{0}(2M-M^{\prime})l-B^{2}J]=
-\frac{l}{2}\oint d\phi[\xi^{-}M^{+}+\xi^{+}M^{-}],\label{eq:charge}
\end{equation}
where $M^{\pm}\equiv 2M-M^{\prime}\pm\frac{J}{l}$ and

$\xi^{\pm}\equiv -(B^{0}\pm B^{2})=\xi^{\pm}(t^{\pm})$
with $t^{\pm}\equiv\frac{t}{l}\pm\phi$.

Owing to this charge, we can compute the Poisson bracket of two
generators following the usual definition
(see, e.g., ref.\cite{banados}). The result is
\begin{equation}
\{G[\xi],G[\eta]\}_{P.B.}=-G[[\xi,\eta]+\delta_{\xi}\eta-\delta_{\eta}
\xi+(\cdots)]
+2\oint_{\partial\Sigma}{\rm Tr}[\xi\delta_{\eta}(\tilde{A})_{0}],
\end{equation}
where $\delta_{\eta}\xi=\{\xi,G[\eta]\}_{P.B.}$, and $(\cdots)$ denotes
a linear combination of the constraints. The last term in the
R.H.S. gives the central term.
If we solve the constraints and formally take the Dirac bracket,
we find  that the charges form
a pseudo-conformal algebra with a central term\cite{brown}:
\begin{equation}
\{Q[\xi],Q[\eta]\}_{D.B.}=-Q\bigl[[\xi,\eta]\bigr]+l\oint
d\phi(-\xi^{+}\partial_{\phi}^{3}\eta^{+}+\xi^{-}\partial_{\phi}^{3}
\eta^{-}),
\label{eq:DBC}
\end{equation}
where we have given the central term explicitly. $\delta_{\xi}\eta$
and $\delta_{\eta}\xi$ do not contribute to the R.H.S., because only
the $O(r)$ terms in eq.(\ref{eq:asgauge}) contribute to the
expression of the charge (\ref{eq:charge}) and because
$\delta_{\xi}\eta$
is of $O(\frac{1}{r})$.

By substituting
$$([\xi,\eta])^{\pm}=\pm(\xi^{\pm}\partial_{\phi}\eta^{\pm}-
\eta^{\pm}\partial_{\phi}\xi^{\pm})$$
into (\ref{eq:DBC}), we can extract the Dirac bracket of $M^{\pm}$,
\begin{eqnarray}
\{M^{\pm}(\phi),M^{\pm}(\phi^{\prime})\}_{D.B.}&=&
\mp\frac{2}{l}[\partial_{\phi}\delta(\phi,\phi^{\prime})M^{\pm}
(\phi^{\prime})-\partial_{\phi^{\prime}}\delta(\phi,\phi^{\prime})
M^{\pm}(\phi)+2\partial_{\phi^{\prime}}^{3}\delta(\phi,\phi^{\prime})],
\nonumber \\*
\{M^{+}(\phi),M^{-}(\phi^{\prime})\}_{D.B.}&=&0.\label{eq:DBM}
\end{eqnarray}

While in principle transformation of $M^{\pm}$ under the asymptotic
gauge transformation (\ref{eq:asgauge}) can be computed by examining
the asymptotic form of $\delta_{\xi}A$ up to $O(\frac{1}{r})$,
it is much easier
to use the charge (\ref{eq:charge}) and the Dirac
bracket(\ref{eq:DBM}).
We will give the result only:
\begin{equation}
\delta_{\xi}M^{\pm}=\{M^{\pm},Q[\xi]\}_{D.B.}=
\pm(\xi^{\mp}\partial_{\phi}M^{\pm}+2\partial_{\phi}\xi^{\mp}M^{\pm}
-2\partial_{\phi}^{3}\xi^{\mp}). \label{eq:virasoro}
\end{equation}
Because the Hamiltonian $B(\partial\Sigma)$
given by eq.(\ref{eq:hamiltonian})is a
particular charge $Q[\xi]$ with $\xi^{+}=\xi^{-}=-1/l$,
we find the time evolution of $M^{\pm}$ as
$$\partial_{t}M^{\pm}=\{M^{\pm},B(\partial\Sigma)\}_{D.B.}
=\mp\frac{1}{l}\partial_{\phi}M^{\pm},$$
After solving the equations of motion(\ref{eq:EOM}),
$M^{\pm}$ thus reduces to the function defined on $S^{1}$
(coordinatized by $t^{\mp}$),
\begin{equation}
M^{\pm}(t,\phi)=M^{\pm}(t^{\mp}).\label{eq:teich}
\end{equation}
By noticing that $\xi^{+}$ and $\xi^{-}$ are respectively functions
of $t^{+}$ and of $t^{-}$ only, transformation(\ref{eq:virasoro})
turns out to be the centrally extended transformation of  a quadratic
differential on $S^{1}$. The integrated version of (\ref{eq:virasoro})
is as follows
\footnote{$\{,\}$ denotes the Schwartzian derivative which is defined
to be
$$
\{\zeta,z\}\equiv\frac{d^{3}\zeta/dz^{3}}{d\zeta/dz}-\frac{3}{2}
\left(\frac{d^{2}\zeta/dz^{2}}{d\zeta/dz}\right)^{2}.
$$}
\begin{equation}
M^{\pm}(t^{\mp})\rightarrow M^{\pm\prime}(t^{\mp})=
\left(\frac{dt^{\mp\prime}}{dt^{\mp}}\right)^{2}M^{\pm}(t^{\mp\prime})
-2\{t^{\mp\prime},t^{\mp}\},\label{eq:Virasoro}
\end{equation}
where $t^{\pm\prime}(t^{\pm})$ is a \lq\lq new coordinate on $S^{1}$"
which is subject to the periodic condition:
$$t^{\pm\prime}(t^{\pm}+2\pi)=t^{\pm\prime}(t^{\pm})+2\pi.$$

The effective theory of the asymptotic $ADS^{3}$ spacetimes thus
reduces
to a pseudo-conformal field theory defined on a cylinder which is
coordinatized by $(t,\phi)$ and whose metric is conformal to
$$
ds_{(2)}^{2}=-\frac{dt^{2}}{l^{2}}+d\phi^{2}.
$$
$M^{\pm}$ plays the role of \lq\lq stress-energy tensor"
in this effective theory.

Because the asymptotic gauge transformation(\ref{eq:asgauge}) induces
the diffeomorphism of the metric (in the case where the metric
is nondegenerate\cite{witte}), we naively expect that the two sets of
parameters
$(M^{+},M^{-})$ and $(M^{\prime+},M^{\prime-})$ give diffeomorphism
equivalent
spacetimes when they are related with each other by
eq.(\ref{eq:Virasoro}). To see whether this is indeed the case
requires a detailed investigation on the global structure
such as singularity, horizon, etc..
This statement is , however, true as long as the asymptotic behavior of
the spacetime is concerned. The moduli space of asymptotically
$ADS^{3}$
spacetimes is therefore given by
$${\cal Q}^{+}\times{\cal Q}^{-},$$
where ${\cal Q}^{\pm}$ is the space of centrally extended quadratic
differentials on $S^{1}$ with periodic coordinates $t^{\pm}$.

(Remark:) In order to check that the asymptotic gauge transformation
(\ref{eq:asgauge}) preserves the asymptotic form (\ref{eq:asp}),
we have to give next-to-leading order terms in (\ref{eq:asp}) so that
the curvature should vanish asymptotically up to eq.(\ref{eq:eom}),
and we have to give adequate $O(\frac{1}{r^{2}})$ terms in
eq.(\ref{eq:asgauge}). The next-to-leading order terms are determined
uniquely by six arbitrary functions of $(t,\phi)$, which gives (leading
order terms of) the gauge degrees of freedom irrelevant to the
asymptotic
physics. Invariance of the action (\ref{eq:EPac}) under asymptotic
transformation (\ref{eq:asgauge})
is established only after we give the next-to-leading order terms in
(\ref{eq:asp}).


\section{Generic Solution}

Now we explicitly solve equations of motion (\ref{eq:EOM}) to obtain
the spacetime which is parametrized by arbitrary
\lq\lq stress energy tensor" $M^{\pm}(t^{\mp})$.
To eliminate the gauge degrees of freedom which are at most of
$O(\frac{1}{r^{2}})$ and which are irrelevant to the asymptotic
physics,
we have to impose appropriate
gauge-fixing conditions which have nonvanishing Poisson brackets
with the first class constraints.
First we fix the local Lorentz gauge degrees of freedom.
This can be done by fixing the triad parts. We impose
\begin{equation}
e^{0}_{r}=e^{0}_{\phi}=e^{2}_{r}=0.\label{eq:lLgauge}
\end{equation}
In order to fix the remaining gauge, we have to set further three
gauge-fixing conditions, two of which we will take
\footnote{The remaining condition is given by $e^{2}_{\phi}=r$ or by
$e^{1}_{r}=\frac{l}{r}$.
Our gauge seems to be good as least in the sense that
the matrix of Poisson brackets between the constraints and the
gauge-fixing
conditions does not degenerate weekly almost everywhere sufficiently
inside the infinity. At infinity, however, the matrix asymptotically
degenerates. This seems to be the source of the appearance of the
asymptotic gauge degrees of freedom.}
\begin{equation}
\omega^{01}_{r}=\omega^{12}_{r}=0. \label{eq:diffgauge}
\end{equation}
{}{}

To simplify the analysis we will make the ansats
that the metric components can be expressed by power series
$\sum_{n\geq n_{0}}a_{n}r^{-n}$ with $a_{n}$ being in general
some function of $(t,\phi)$. Owing to
this ansats, we can assert that:
\begin{equation}\begin{array}{c}
\mbox{\it if a quantity $q$ of $O(1/r^{n})$ with $n>0$ satisfies } \\
\mbox{$\partial_{r}q=0$, then $q=0$.}
\end{array}\label{eq:ansatz}
\end{equation}

By properly combining eqs.(\ref{eq:EOM}),
(\ref{eq:lLgauge}-\ref{eq:ansatz}), we find
\begin{eqnarray}
e^{1}&=&ld\rho\qquad\quad\mbox{\footnotesize $\rho=\ln\frac{r}{l}-
\frac{M^{\prime}l^{2}}{4r^{2}}+O(\frac{1}{r^{3}})$}, \nonumber \\*
\omega^{20}&=&d\sigma\qquad\quad\mbox{\footnotesize $\sigma
=-\frac{Jl}{4r^{2}}+O(\frac{1}{r^{3}})$}. \label{eq:rhosig}
\end{eqnarray}

Let us now consider the gauge transformated
connection:
\begin{equation}
A^{\prime}\equiv gAg^{-1}-dgg^{-1}\quad ,
g\equiv\exp(P_{1}l\rho+J_{1}\sigma).
\end{equation}
The result of substituting eqs.(\ref{eq:asp})(\ref{eq:lLgauge}-
\ref{eq:rhosig})is considerably simplified as follows:
\begin{equation}\begin{array}{c}
A^{\prime}=P_{0}l[(1-\frac{2M-M^{\prime}}{4}+O(\frac{1}{r}))\frac{dt}
{l}
+(\frac{J}{4l}+O(\frac{1}{r}))d\phi]\\
+P_{2}l[(1+\frac{2M-M^{\prime}}{4}+O(\frac{1}{r}))d\phi
+(-\frac{J}{4l}+O(\frac{1}{r}))\frac{dt}{l}] \\
+J_{0}[(1-\frac{2M-M^{\prime}}{4}+O(\frac{1}{r}))d\phi
+(\frac{J}{4l}+O(\frac{1}{r}))\frac{dt}{l}]\quad\\
+J_{2}[(1+\frac{2M-M^{\prime}}{4}+O(\frac{1}{r}))\frac{dt}{l}
+(-\frac{J}{4l}+O(\frac{1}{r}))d\phi].
\end{array}
\end{equation}
By solving the equations of motion $dA^{\prime}+A^{\prime}\wedge
A^{\prime}=0$ and by
using (\ref{eq:ansatz}), we see that  all the $O(1/r)$-terms
in $A^{\prime}$ vanish and that the following equations hold:
\begin{equation}
\partial_{\phi}(2M-M^{\prime})+\partial_{t}J=0\quad,
l^{2}\partial_{t}(2M-M^{\prime})+\partial_{\phi}J=0.\label{eq:eom}
\end{equation}
We can easily solve these equations and find
\begin{equation}
M^{\pm}\equiv 2M-M^{\prime}\pm\frac{J}{l}=M^{\pm}(t^{\mp}),
\end{equation}
where $t^{\pm}\equiv\frac{t}{l}\pm\phi$.  This is the very equation
(\ref{eq:teich}) which has been obtained by the general analysis.

To obtain the explicit form of  the SO$(2,2)$ connection $A$, we have
only to perform the inverse transformation
$$A=g^{-1}A^{\prime}g-dg^{-1}g. $$
The result is:
\begin{eqnarray}
A&=&P_{0}lB\frac{dt}{l}+P_{2}l(Cd\phi+D\frac{dt}{l})+P_{1}ld\rho
\nonumber \\*
&
&+J_{0}Bd\phi+J_{2}(C\frac{dt}{l}+Dd\phi)+J_{1}d\sigma,\label{eq:Gsol}
\end{eqnarray}
with
\begin{equation}\left\{\begin{array}{llc}
B&=&C^{-1}[e^{2\rho}-\frac{M^{+}M^{-}}{16}e^{-2\rho}]\\
C&=&[e^{2\rho}+\frac{M^{+}+M^{-}}{4}+
\frac{M^{+}M^{-}}{16}e^{-2\rho}]^{1/2} \\
D&=&-J/2lC\\
e^{2\sigma}&=&
[e^{2\rho}+\frac{M^{+}}{4}]^{-1}[e^{2\rho}+\frac{M^{-}}{4}].
\end{array}\right. \label{eq:coeffs}
\end{equation}
The last equation is necessary for $A$ to satisfy the gauge-fixing
condition $e^{0}_{\phi}=0$. This connection gives the spacetime which
is specified by the numerical values of the \lq\lq stress-energy
tensor"
$M^{\pm}$.

To look for the residual gauge degrees of freedom, it is convenient to
use (anti-)chiral SL$(2,{\bf R})$ connections $A^{(\pm)}\equiv
J_{a}(\frac{1}{2}\epsilon^{a}_{\mbox{ }bc}\omega^{bc}\pm e^{a}/l)$:
\begin{equation}
A^{\prime(\pm)}=\pm J_{0}(1-\frac{M^{\mp}}{4})dt^{\pm}
+J_{2}(1+\frac{M^{\mp}}{4})dt^{\pm}.\label{eq:chiral}
\end{equation}
The gauge transformations which keeps this form of $A^{\prime(\pm)}$
are uniquely determined up to one arbitrary function
$\xi^{\pm}(t^{\pm})$:
\begin{equation}
\xi^{\prime(\pm)}=\mp J_{0}\{\xi^{\pm}-\frac{1}{2}(\frac{M^{\mp}}{2}
-\partial_{\pm}^{2})\xi^{\pm}\}\pm J_{1}\partial_{\pm}\xi^{\pm}
-J_{2}\{\xi^{\pm}+\frac{1}{2}(\frac{M^{\mp}}{2}
-\partial_{\pm}^{2})\xi^{\pm}\}. \label{eq:chig}
\end{equation}
The induced transformation of $M^{\pm}$ is given by
\begin{equation}
\delta_{\xi}M^{\pm}=-\xi^{\mp}\partial_{\mp}M^{\pm}-2\partial_{\mp}
\xi^{\mp}M^{\pm}+2\partial_{\mp}^{3}\xi^{\mp}.
\end{equation}

This is nothing but the transformation(\ref{eq:virasoro}). So we
realize
that, in our space of solutions, there are no gauge degrees of freedom
other than the transformation of centrally extended quadratic
differentials
on $S^{1}$, i.e., transformation (\ref{eq:virasoro}).

In passing, the asymptotic gauge transformation (\ref{eq:asgauge}) in
our
gauge is
recovered by first setting $\xi^{\pm}=-(B^{0}\pm B^{2})$ and then
substituting eq.(\ref{eq:coeffs}) into the following expression of
$\xi$
\footnote{This form of $\xi$ is derived by considering

the following sequence of transformations:
$$
A\rightarrow A^{\prime}\rightarrow
A^{\prime}+\delta_{\xi^{\prime}}A^{\prime}
\rightarrow A+\delta_{\xi}A.
$$}
\begin{equation}
\xi=g^{-1}\xi^{\prime}g+J_{1}\delta_{\xi}\sigma+P_{1}l\delta_{\xi}\rho.
\end{equation}


\section{Simple Examples}

To investigate spacetimes given by(\ref{eq:Gsol}), it is nessesary
to choose an adequate radial coordinate.
In order to see that our solution involves 3DBH (possibly with a
negative
mass), we first fix the gauge by
\begin{equation}
e^{2}_{\phi}=\tilde{r}. \label{eq:rgauge}
\end{equation}
The metric constructed from (\ref{eq:Gsol}) is then expressed as
\footnote{In the last two sections we have seen that the combination
$2M-M^{\prime}$
plays a crucial role. We will henceforth rename $2M-M^{\prime}$ as
$M$ because of the notational convenience.}
\begin{equation}\begin{array}{l}
ds^{2}=-(\frac{\tilde{r}^{2}}{l^{2}}-M+\frac{J^{2}}{4\tilde{r}^{2}})dt^
{2}+
(\tilde{r}d\phi-\frac{J}{2\tilde{r}}dt)^{2}\\
+\frac{1}{(\frac{\tilde{r}^{2}}{l^{2}})^{2}-M\frac{\tilde{r}^{2}}{l^
{2}}+
\frac{J^{2}}{4l^{2}}}\left\{\frac{\tilde{r}d\tilde{r}}{l}-\frac{ldM}{4}
-
\frac{ld(M^{+}M^{-})}{4M^{+}M^{-}}(\frac{\tilde{r}^{2}}{l^{2}}-\frac{M}
{2}-
\sqrt{(\frac{\tilde{r}^{2}}{l^{2}})^{2}-M\frac{\tilde{r}^{2}}{l^{2}}+
\frac{J^{2}}{4l^{2}}})\right\}^{2}.
\end{array}\label{eq:BTZ}
\end{equation}
When $M^{+}$ and $M^{-}$ are both constant, this gives 3DBH\cite{BTZ}
possibly with a negative mass. When not both of $M^{+}$ and $M^{-}$ are
constant and $M^{+}M^{-}>0$, however, this metric becomes essentially
complex in the region $r_{-}<\tilde{r}<r_{+}$ with
$\frac{r_{\pm}^{2}}{l^{2}}\equiv
\frac{M\pm\sqrt{M^{+}M^{-}}}{2}$.
It would be natural to consider that such an eccentric metric is
physically
meaningless unless we can bypass the region of complex metric e.g.
by a coordinate redefinition.

Thus we take the following radial gauge
\begin{equation}
e^{2\rho}=\frac{r^{2}}{l^{2}}\equiv\zeta\quad
(i.e.,\quad e^{1}_{r}=\frac{l}{r}).
\label{eq:new}
\end{equation}
The metric in this gauge is
\begin{eqnarray}
ds^{2}&=&-\frac{(\zeta^{2}-\frac{M^{+}M^{-}}{16})^{2}}
{(\zeta+\frac{M^{+}}{4})(\zeta+\frac{M^{-}}l{4})\zeta}dt^{2}
+\frac{l^{2}}{4\zeta^{2}}d\zeta^{2}\nonumber \\*
& &+\left(\sqrt{\frac{(\zeta+\frac{M^{+}}{4})(\zeta+\frac{M^{-}}{4}}
{\zeta}}ld\phi-\frac{J}{2l}
\sqrt{\frac{\zeta}{(\zeta+\frac{M^{+}}{4})
(\zeta+\frac{M^{-}}{4}}}dt\right)^{2}\label{eq:NEW} \\
&=&-(\frac{r^{2}}{l^{2}}-\frac{M}{2}+\frac{M^{+}M^{-}l^{2}}{16r^{2}})dt
^{2}+
(r^{2}+\frac{Ml^{2}}{2}+\frac{M^{+}M^{-}l^{4}}{16r^{2}})d\phi^{2}
-Jdtd\phi+\frac{l^{2}}{r^{2}}dr^{2}. \nonumber
\end{eqnarray}
This metric is obviously real and has the signature of Lorentzian
spacetime throughout the region of real $\zeta$ (except
$\zeta=0,-\frac{M^{\pm}}{4},\frac{\sqrt{M^{+}M^{-}}}{4}$ surfaces).
So we regard this new radial gauge (\ref{eq:new}) as a natural
choice of the radial coordinate.

To see what happens when we change the radial coordinate from
(\ref{eq:rgauge}) to (\ref{eq:new}), we depict in fig.1 the behavior
of conventional radial coordinate $\tilde{r}$
in the complex $\zeta$-plane.
We can see that the region $r_{-}<\tilde{r}<r_{+}$ of complex metric
draws a semicircle in the complex $\zeta$-plane.
This region of complex metric corresponds to the
region between the outer horizon and the inner horizon when
$(M^{+},M^{-})=const.$.
It therefore seems more natural to remove the region inside
the outer horizon at least when we consider the 3DBH as
belonging to our general solution.

Next we briefly investigate the spacetime
which is described by the metric (\ref{eq:NEW}) with $(M^{+},M^{-})
=const.$.
There are three cases depending on the signature of
$(M^{+},M^{-})$.

i)$M^{+}\geq 0,M^{-}\geq 0$.

In this case the spacetime is a three dimensional
black hole. The parametrization  in the (2+2)-dimensional Minkowskii
space
is given by ref.\cite{BTZ}:
\footnote{We consider that
$$T^{2}-X^{2}-Y^{2}+Z^{2}=L^{2}$$
holds. The metric is obtained by substituting the
parametrization into the pseudo-Minkowski metric:
$$
ds^{2}=-dT^{2}+dX^{2}+dY^{2}-dZ^{2}.
$$}
\begin{eqnarray}
(T,X)=\frac{r^{2}+\frac{\sqrt{M^{+}M^{-}}}{4}l^{2}}
{(M^{+}M^{-})^{1/4}r}(\cosh\tilde{\phi},\sinh\tilde{\phi})
\quad\mbox{\footnotesize
$\tilde{\phi}\equiv\frac{-\sqrt{M^{-}}t^{+}+\sqrt{M^{+}}t^{-}}{2}$},
\nonumber \\*
(Y,Z)=\frac{r^{2}-\frac{\sqrt{M^{+}M^{-}}}{4}l^{2}}
{(M^{+}M^{-})^{1/4}r}(\cosh\tilde{t},\sinh\tilde{t})
\quad\mbox{\footnotesize
$\tilde{t}\equiv\frac{\sqrt{M^{-}}t^{+}+\sqrt{M^{+}}t^{-}}{2}$}.
\label{eq:PP}
\end{eqnarray}
At $r=r_{0}\equiv\frac{(M^{+}M^{-})^{1/4}l}{2}$ there exists a horizon
of the Rindler space type which splits the spacetime into two
causally-independent regions $r>r_{0}(Y>|Z|)$ and $r<r_{0}
(Y<-|Z|)$. This horizon $r=r_{0}$ is the remnant of the outer
horizon of  3DBH. $r=\infty$ and $r=0$ correspond
respectively to spatial infinities of the two regions.\\
ii)$M^{+}<0,M^{-}<0$.
The spacetime is a \lq\lq negative-mass black hole".
The conical and helical singularity appears at
$r=r_{0}\equiv\frac{(M^{+}M^{-})^{1/4}l}{2}$.This spacetime
involves \lq\lq obvious"
closed timelike curves (CTC)with $t$ and $r$ being constant, in the
region
$r_{0}<r<{\rm max}\{\frac{(-M^{+})^{1/2}l}{2},
\frac{(-M^{+})^{1/2}l}{2}\}$.  Hence this case is usually ruled out
from the physical
spectrum\cite{hawk}
(except the case with $M^{+}=M^{-}$, where the CTC's necessarily pass
through
the conical singularity at $r=r_{0}$).\\
iii)$M^{+}M^{-}<0$.This spacetime does not have either conical
singularity or
horizon and so we can naturally take the domain of $r$ to be
$(0,\infty)$. This spacetime, however, necessarily involves
the obvious CTC's in the region  $r<{\rm
max}\{\frac{(-M^{+})^{1/2}l}{2},
\frac{(-M^{+})^{1/2}l}{2}\}$
and so we usually exclude this from the physical spectrum.

Now, in order to look for nontrivial spacetimes, let us investigate
the case with $M^{\pm}=m^{\pm}_{0}+\delta M^{\pm}$, where
$m^{\pm}_{0}$ is a constant and $\delta M^{\pm}$ is a small
fluctuation.
For $M^{\pm}$ not to be gauge equivalent to $m^{\pm}_{0}$, we must
have $\delta M^{\pm}$ which cannot be absorbed into the gauge
transformations, i.e.

\begin{equation}
\delta M^{\pm}\neq -\xi^{\mp}\partial_{\mp}m^{\pm}_{0}-2\partial
_{\mp}\xi^{\mp}m^{\pm}+2\partial_{\mp}^{3}\xi^{\mp}
\end{equation}
for any well-defined function $\xi^{\mp}(t^{\mp})$ on $S^{1}$.
By substituting the fourier decomposition $\xi^{\pm}=\sum\xi^{\pm}_{n}
e^{int^{\pm}}$, we find the two cases.

a) When $m^{\pm}_{0}\neq -n^{2}$
for $^{\forall}n\in{\bf Z}\backslash\{0\}$, only $\delta
M^{\pm}=const.$
survives.  Thus we have only to consider the constant $M^{\pm}$
which gives 3DBH.

b)When $m^{\pm}_{0}=-n^{2}$ with $^{\exists}n\in{\bf
Z}\backslash\{0\}$,
$\sin nt^{\mp}$ and $\cos nt^{\mp}$ also survive as nontrivial $\delta
M^{\pm}$. We therefore expect the appearance of nontrivial
spacetimes in this case.

We will only consider the case with $m^{+}_{0}=m^{-}_{0}=-n^{2}$.

Otherwise case b) corresponds to the fluctuation about physically
irrelevant spacetimes in which naked CTC's appear.
To see the behavior of new solutions, it is sufficient to investigate
the case $m^{+}_{0}=m^{-}_{0}=-1$.  We will only consider
the contribution of oscillating fluctuation. By a proper
constant shift of $(t,\phi)$, we can take the form of $M^{\pm}$ as
\begin{equation}
M^{\pm}=-1+4\epsilon^{\pm}\cos t^{\mp},
\end{equation}
where $\epsilon^{\pm}$ is a positive infinitesimal constant.
Substituting this into (\ref{eq:NEW}), we find the metric of the new
spacetime:
\begin{equation}\begin{array}{ll}
ds^{2}=&-\frac{(\zeta+\frac{1}{4}-\epsilon^{+}\cos t^{-})
(\zeta+\frac{1}{4}-\epsilon^{-}\cos t^{+})}{\zeta}dt^{2}+
l^{2}\frac{(\zeta-\frac{1}{4}+\epsilon^{+}\cos t^{-})
(\zeta-\frac{1}{4}+\epsilon^{-}\cos t^{+})}{\zeta}d\phi^{2}\\
&-2l(\epsilon^{+}\cos t^{-}-\epsilon^{+}\cos t^{+})dtd\phi
+l^{2}\frac{d\zeta^{2}}{4\zeta^{2}}.
\end{array}\label{eq:ours}
\end{equation}
This spacetime would not probably have any curvature singularity
because it is a solution of Einstein's equations in (2+1)-dimensions,
which give at  most conical singularities.
Because of the fluctuation it is difficult to see whether
there exist conical singularities in this spacetime.

We will  only investigate whether CTC's exist or not.
Because CTC's pass through the region $g_{\phi\phi}<0$ at least twice,
it suffices to consider CTC's of the following form
\begin{equation}
x(\phi)=(t,\phi,\zeta)=(t_{0}+
\delta t(\phi),\phi,\frac{1}{4}+\delta\zeta(\phi)),\label{eq:ctc}
\end{equation}
where
$t_{0}$ is a constant and $\delta t$ and $\delta\zeta$ are small
fluctuations of $O(\epsilon)$ which are periodic in $\phi$. The
condition
for the $x(\phi)$ to be a CTC is for all $\phi$ the following
inequality holds:
{\footnotesize$$
0\geq(\frac{ds}{d\phi})^{2}=-(\frac{d(\delta t)}{d\phi}+l\epsilon^{+}
\cos t^{-}_{0}-l\epsilon^{-}\cos t^{+}_{0})^{2}+l^{2}
(2\delta\zeta+\epsilon^{+}\cos t^{-}_{0}+\epsilon^{-}\cos t^{+})^{2}
+4l^{2}(\frac{d(\delta\zeta)}{d\phi})^{2}+O(\epsilon^{3}),
$$}
where we have set $t^{\pm}_{0}=\frac{t_{0}}{l}\pm\phi$.
We may restrict the analysis to  the case where
$$\delta t=A\cos\phi+B\sin\phi,\quad
\delta\zeta=C\cos\phi+D\sin\phi.
$$
Substituting this into the above inequality and making an elementary
but
tedious analysis, we find that CTC's of the form
(\ref{eq:ctc}) appear in the region
{\footnotesize$$
\frac{1}{4}-\frac{1}{2}\sqrt{(\epsilon^{+}+\epsilon^{-})^{2}\cos^{2}
\frac{t_{0}}{l}+(\epsilon^{+}-\epsilon^{-})^{2}\sin^{2}\frac{t_{0}}{l}}
\leq\zeta\leq\frac{1}{4}+\frac{1}{2}\sqrt{(\epsilon^{+}+\epsilon^{-})^
{2}
\cos^{2}\frac{t_{0}}{l}
+(\epsilon^{+}-\epsilon^{-})^{2}\sin^{2}\frac{t_{0}}{l}}
$$}
It would therefore be probable to exclude the new spacetime
(\ref{eq:ours})
from the physical spectrum, unless the CTC's can be shielded by some
singular structure.


\section{Several Aspects of the Moduli Space}

In this section
we investigate some properties of  the moduli space, which
is a direct product of two copies of the space ${\cal Q}$ of
centrally extended quadratic differentials $T(\phi)$ on $S^{1}$.
The transformation of $T$ generated by a vector field $v(\phi)
\frac{\partial}{\partial\phi}$ is already given by (\ref{eq:virasoro}):
\begin{equation}
\delta_{v}T=-vT^{\prime}-2v^{\prime}T+2v^{\prime\prime\prime},
\label{eq:virasoro2}
\end{equation}
where $v^{\prime}\equiv\frac{d}{d\phi}v$.
Thus our problem can be translated into that of finding all the
functions $T(\phi)$ which do not mutually transform by
(\ref{eq:virasoro2}).

Instead of dealing with $T$ itself, it is convenient to consider the
\lq\lq stabilizer" of a given $T$. A stabilizer of $T$ is a vector
field
$f(\phi)\frac{\partial}{\partial\phi}$ which leaves $T$ unchanged:
\begin{equation}
0=\delta_{f}T=-fT^{\prime}-2f^{\prime}T+2f^{\prime\prime\prime}.
\label{eq:stabil}
\end{equation}
{}{}

According to ref.\cite{witte3}, we can say the following.\\
1)For a fixed $T(\phi)$, the stabilizers form a vector space whose
dimension is either 1 or 3.\\
2)If a stabilizer $f$ is given, $T$ can be expressed by using $f$:
\begin{equation}
T=\frac{d-(f^{\prime})^{2}}{f^{2}}+2\frac{f^{\prime\prime}}{f},
\label{eq:stab}
\end{equation}
where $d$ is a constant adjusted by requiring the regularity of $T$.
\footnote{Under the vector transformation:
$f(\phi)\rightarrow\tilde{f}(\phi)=\frac{d\phi}{d\tilde{\phi}}
f(\tilde{\phi})$, the $T$ given by (\ref{eq:stab}) is subject to the
desired
transformation:
$$
T(\phi)\rightarrow\tilde{T}(\phi)=(\frac{d\tilde{\phi}}{d\phi})^{2}
T(\tilde{\phi})-2\{\tilde{\phi},\phi\}.
$$}\\
3)The stabilizers necessarily belong to one of the following three
types: O) $f$ has no zeros. Then $d$ is an arbitrary parameter.
I) $f$ has only (even number of) single zeros. At each zero
$(f^{\prime})^{2}$ has the same value which should equal to $d$.
II) $f$ has only double zeros where $f^{\prime\prime\prime}$ must
vanish. $d$ is zero in this case.\\
4) The type O)-stabilizers are diffeomrphic to $f=const.$.
Because $\frac{d\phi}{f(\phi)}$ is then nonsingular and, by taking an
appropriate
coordinate $\tilde{\phi}$, we can set:
\begin{equation}
\frac{d\phi}{f(\phi)}=\frac{d\tilde{\phi}}{a},\qquad\left(\frac{2\pi}
{a}\equiv
\oint\frac{d\phi}{f(\phi)}\right).
\end{equation}
By substituting $f(\phi)=a\frac{d\phi}{d\tilde{\phi}}$ into
(\ref{eq:stab}),
we can see that the $T(\phi)$ with a type O)-stabilizer is equivalent
to
a constant $T$.\\
5)The type I)-stabilizers are splitted into the non-diffeomorphic
classes
which are characterized by the number of zeros $2n$ and the magnitude
of
$\Delta$ which is defined by
\begin{equation}
\Delta=\lim_{\epsilon\rightarrow 0}\int_{|\phi-\phi_{k}|>\epsilon}d\phi
\frac{|f^{\prime}(\phi_{k})|}{f(\phi)},
\end{equation}
where $\phi_{k}$ ($k=1,\cdots,2n$) are the zeros of $f$.\\
6)Two type II)-stabilizers are diffeomorphic with each other
if they have the same
number of double zeros and the same signature of $U(f)$.
\footnote{$U(f)$ is defined as follows. In the interval $\phi_{k}<\phi
<\phi_{k+1}$, we consider $\tau_{k}$which is defined by
$$d\tau_{k}=\frac{d\phi}{f(\phi)},$$ and which is normalized by:
$\tau_{k}\sim\frac{-1}{\phi-\phi_{k}}\quad as\quad
\phi\rightarrow\phi_{k}+0.\quad$Then $\tau_{k}$ always behaves as
$\tau_{k}\sim\frac{-1}{\phi-\phi_{k+1}}+a_{k},$
for $\phi\rightarrow\phi_{k+1}-0$. The definition of $U(f)$ is:
$U(f)=\sum_{k}a_{k}.$}

We can construct concrete realization of generical type I)- and
type II)-stabilizers.

The type I)-stabilizers are represented by
\begin{equation}
f_{n,a}(\phi)=\sin n\phi+a\cos2n\phi,\quad-1<a<1.\label{eq:typei}
\end{equation}
This $f_{n,a}$ has $2n$ zeros and
$\Delta=n\pi\frac{4a^{2}-1+\sqrt{1+8a^{2}}}{2a\sqrt{1-a^{2}}}.$
The $T(\phi)$ which is stabilized by $f_{n,a}$ is
\begin{equation}
T(\phi)=-n^{2}\frac{16a^{2}(\sin n\phi-\alpha)^{2}+4a
(\sin n\phi-\alpha)-2+\sqrt{1+8a^{2}}}{4a^{2}(\sin n\phi-\alpha)^{2}},
\label{eq:type1}
\end{equation}
where $\alpha=\frac{1+\sqrt{1+8a^{2}}}{4a}$. From eq.(\ref{eq:typei})
we can see that $f_{n}(\phi)=\sin n\phi$ corresponds to the limit
$a\rightarrow 0$ in which $\Delta\rightarrow 0$ and
$T(\phi)\rightarrow -n^{2}$.

The type II)-stabilizers should be diffeomrphic to
\begin{equation}
\tilde{f}_{n,b}(\phi)=1-(1-b)\cos n\phi-b\cos2n\phi,
\quad-\frac{1}{3}<a<1, \label{eq:typeii}
\end{equation}
which has $n$ double zeros and $U(\tilde{f}_{n,b})=\frac{2\pi n^{2}b}
{\sqrt{(1-b)(1+3b)}}$. This $\tilde{f}_{n,b}$ stabilizes
\begin{equation}
T(\phi)=-n^{2}\frac{1+3b^{2}+(6b+2b^{2})(1+2\cos n\phi)+4b^{2}
(1+2\cos n\phi)^{2}}{(1+b(1+2\cos n\phi))^{2}}.
\label{eq:type2}
\end{equation}
If we set $b=0$, eq.(\ref{eq:typeii}) reduces
to $\tilde{f}_{n}=1-\cos n\phi$ which has $U(\tilde{f}_{n})=0$
and which stabilizes $T(\phi)=-n^{2}$. On account of statement 6),

the space of the type II)-stabilizers splits into three equivalence
classes under diffeomorphisms on $S^{1}$, whose representatives are
$\tilde{f}_{n,b}$ with $b$ being positive, zero, and negative
respectively.

Let us now consider the \lq\lq tangent space" $T{\cal Q}$
of ${\cal Q}$, i.e.
the space of small fluctuations $\delta T$ which cannot be
absorbed into the transformation
given by (\ref{eq:virasoro2}).
By looking at
eq.(\ref{eq:stab}) and its associated footnote, we see that,
at least when the space of the stabilizers is one-dimensional,
we have only to look for the fluctuation $\delta f$ of the
stabilizer which cannot be expressed by the vector transformation:
\begin{equation}
\delta f=\delta_{v}f=-vf^{\prime}+v^{\prime}f.
\end{equation}
This equation has a formal solution
\begin{equation}
v(\phi)=f\int d\phi\frac{\delta f}{f^{2}}.
\end{equation}
Our problem reduces to that of finding $\delta f$ which
is well-defined on $S^{1}$ and which
gives non-periodic $v$. The desired $\delta f$ is given by
$f^{2}$, which gives $v=f\phi$.
\footnote{Multiplication by a regular function $C(\phi)$ with

$\oint d\phi C(\phi)=2\pi C_{0}\neq 0$ does not influence essential
results. The reasoning is as follows. We can decompose
$C(\phi)$ into $C_{0}+c(\phi)$, where
$C_{0}$ is a constant and $\oint d\phi c(\phi)=0$. The constant
gives the same result and $c(\phi)$ part gives the portion of
$\delta f$ which is well-defined on $S^{1}$.
While $\delta f=1$, $\delta f=f$, etc. give $v$ which is singular,
such fluctuations also make $\delta T$ singular. We can therefore rule
out
these fluctuations.}
Taking the fluctuation of eq.(\ref{eq:stab}) and substituting
$\delta f=-\frac{1}{2}f^{2}$, we can compute $\delta T$ which cannot be
represented by (\ref{eq:virasoro2}). The result is:
\begin{equation}
\delta T=fT-3f^{\prime\prime}.\label{eq:tangent}
\end{equation}
Because we can find one $\delta T$ in the above form per one
stabilizer $f$, we conclude that the dimension of $T{\cal Q}$ equals
one when there is only one linearly-independent stabilizer.

What about the case where the space of stabilizers being three
dimensions?
By using eq.(\ref{eq:stabil})
we can show that the space of the stabilizers of a given $T$ forms
a Lie algebra under the Lie derivative $[f,g]=fg^{\prime}-gf^{\prime}$.
Because this Lie algebra generates a three dimensional subgroup of
$DiffS^{1}$(the group of diffeomorphisms of a circle,i.e.,
the Virasoro group), it would be natural to identify this group with
$SL^{(n)}(2,{\bf R})$ (the n-fold covering of $SL(2,{\bf R})$). By
performing an appropriate diffeomorphism we can choose the generators
of $SL^{(n)}(2,{\bf R})$ to be $(1,\cos n\phi,\sin n\phi)$, which are
the stabilizers of $T(\phi)=-n^{2}$. Thus $T(\phi)$ which has three
linearly independent stabilizers turns out to be equivalent to
$T=-n^{2}$. Under the action of $SL^{(n)}(2,{\bf R})\subset DiffS^{1}$
the stabilizers
$$f=z+x\cos n\phi+y\sin n\phi$$
of $T=-n^{2}$ split into the equivalence classes whose representatives
are
$f=z$ with $z\in{\bf R}$, $f=y\tilde{f}_{n}$ with $y\in[0,\infty)$,
and $f=\pm f_{n}$ respectively.
The net tangent space of ${\cal Q}$ at a \lq\lq triple stabilizer
point" is thus given by a T-shaped space plus two points (fig.2a).

By gluing the adjacent tangent spaces into together we find the
topology of the whole space ${\cal Q}$ of the centrally extended
quadratic differentials: ${\cal Q}$ is an almost one-dimensional
pectinated space which is constructed by
gluing infinitely-many half-lines ($f_{n,a}$ with fixed $n$) to
one line of ${\bf R}^{1}$ ($f=const.$) at the points $-n^{2}$,
and then by associating two points ($\tilde{f}_{n,\pm\epsilon}$ with
$n$ fixed) to the points $-n^{2}$ (fig.2b).

Finally we mention the relation between ${\cal Q}$ and
the moduli space ${\cal C}$ of

flat  $\tilde{SL}(2,{\bf R})$ connections on
a cylinder, where $\tilde{SL}(2,{\bf R})$ is the universal covering
group of $SL(2,{\bf R})$. It is well known that the moduli space of
flat
connections is parametrized by the conjugation classes of the
holonomies around noncontractible loops\cite{nelson}. Using this
as in the case of the flat $\tilde{SL}(2,{\bf R})$ connections
on a torus\cite{ezawa}, we find that the moduli space
${\cal C}$ is represented by the sum of infinitely many sectors:
\begin{equation}
{\cal C}={\cal C}_{T}\cup\left(\bigcup_{n\in{\bf Z}}{\cal
C}_{S}^{n}\right)
\cup\left(\bigcup_{n\in{\bf Z},\sigma=\pm}{\cal
C}_{N,\sigma}^{n}\right).
\end{equation}
$\tilde{SL}(2,{\bf R})$
connections which parametrize each sector are given, for example, by:
\begin{eqnarray}
{\cal C}_{T}&:&A=J_{0}\beta d\phi,\nonumber \\*
{\cal C}_{S}^{n}&:&A=J_{0}nd\phi
+(J_{2}\cos n\phi+J_{1}\sin n\phi)\tilde{\beta}d\phi,\nonumber \\*
{\cal C}_{N,\pm}^{n}&:&A=J_{0}nd\phi
\pm e^{\lambda}[J_{0}+(J_{2}\cos n\phi+J_{1}\sin n\phi)]d\phi,
\end{eqnarray}
where $\beta\in{\bf R}$ and $\tilde{\beta}\in[0,\infty)$. $\lambda$ is
an
arbitrary parameter which can be absorbed by a boost. We can easily
see that the point $\tilde{\beta}=0\in{\cal C}_{S}^{n}$ coincides with
$\beta=n\in{\cal C}_{T}$, and that ${\cal C}_{N,\pm}^{n}$ are
\lq\lq very close to" (but not connected to)
the point $\beta=n\in{\cal C}_{T}$
(fig.3). From this we find that ${\cal Q}$ is homeomorphic to the
following
subspace of ${\cal C}$:
\begin{equation}
{\cal Q}\approx({\cal C}_{T}\backslash\{0\})/{\bf Z}_{2}
\cup{\cal C}_{N,+}^{0}\cup{\cal C}_{S}^{0}
\cup\left(\bigcup_{n\in{\bf N}}{\cal C}_{S}^{n}\right)
\cup\left(\bigcup_{n\in{\bf N},\sigma=\pm}{\cal
C}_{N,\sigma}^{n}\right),
\label{eq:QandC}
\end{equation}
where ${\bf Z}_{2}$ is generated by the inversion: $A\rightarrow -A$.

This can partially be expected from eq.(\ref{eq:chiral}). When
$M^{\pm}$
is constant, $A^{\prime(\pm)}$ in (\ref{eq:chiral}) can be conjugated
into
the form of $A\in({\cal C}_{T}\backslash\{0\})/{\bf Z}_{2}
\cup{\cal C}_{N,+}^{0}\cup{\cal C}_{S}^{0}$ given above.
The spaces of $M^{\pm}$ each of which is left invariant by a vector
field
$f_{n,a}(t^{\pm})$ and by $\tilde{f}_{n,b}(t^{\pm})$ stretch from the
points
$M^{\pm}=-n^{2}$ (with $n\in{\bf N}$) which correspond to
$A^{(\mp)}=\mp J_{0}ndt^{\mp}$. These spaces should therefore be
related to
${\cal C}_{S}^{n}$ and ${\cal C}_{N,\pm}^{n}$. While it seems to be
difficult
to establish the gauge equivalence between $A\in{\cal C}_{S}^{n}$
(or $A\in{\cal C}_{N,\pm}^{n}$)
and eq.(\ref{eq:chiral}) with $M^{\pm}$ given , for example, by
(\ref{eq:type1}) (or by (\ref{eq:type2})), we conjecture that relation
(\ref{eq:QandC}) is in fact the gauge equivalence relation.

Thus we see that the moduli space of asymptotically
$ADS^{3}$ spacetimes is a subspace of
the moduli space of flat $\tilde{SL}(2,{\bf R})\times\tilde{SL}(2,{\bf
R})$
connections.\footnote{Roughly speaking,
the rest of flat $\tilde{SL}(2,{\bf R})\times\tilde{SL}(2,{\bf R})$
connections
can be interpreted as follows. Because simultaneous changes of
signs in $A^{(\pm)}$ does not affect the metric, the relevant
connections
give spacetimes with opposite orientation. A change of the relative
sign
between $A^{(+)}$ and  $A^{(-)}$ corresponds to exchanging $t$ and
$\phi$. These connections are therefore expected to give spacetime with
CTC's at spatial infinity, which cannot be considered as our universe.}
This result  is expected by naively investigating
Chern-Simons formulation of  (2+1)-anti-de Sitter gravity on a
cylinder.
It is nontrivial, however, that the argument of the chiral connection
$A^{(\pm)}$ is not simply $\phi$ but $\pm t^{\pm}=\pm\frac{t}{l}+\phi$.
This cannot be extracted by a naive analysis of
Chern-Simons formulation.

\section{Discussion}

We have seen that,in (2+1)-dimensions, the moduli space of
asymptotically anti-de Sitter spacetimes modulo diffeomorphisms
is parametrized by two centrally extended quadratic differentials.
We found two new families of solutions corresponding to type I)- and
type II)-stabilizers. From the analysis of  small fluctuations about
$M^{\pm}=-n^{2}$ made in \S 4, however,  these solutions turn out to
have CTC's and thus seem to be ruled out as physically irrelevant.
If we take the equivalence class under all asymptotic transformations,
only 3DBH's and anti-de Sitter space (possibly with a
conical singularity) seem to survive as physically relevant
configurations.

If we consider only the region with sufficiently large radius, however,
these CTC's do not appear. It is possible that we can get rid of
these CTC's by an appropriate surgery of  the spacetime.
Nor we know whether we can really take the equivalence
under all the asymptotic transformations. Behavior of
the metric (\ref{eq:NEW}) would be extremely complicated  when, for
example, $M^{(\pm)}$ has zeros. While we have investigated the
structure
of the moduli space by a somewhat topological consideration, it has not
been shown directly that all the positive-definite $M^{\pm}$ reduce
to a positive constant by an adequate transformation
(\ref{eq:Virasoro}).  For the time being
we only mention that the asymptotic behavior of the
asymptotically $ADS^{3}$ spacetimes
is described by two centrally extended quadratic differentials.
There seem to be numerous issues which require
further investigation.

Finally we comment on the effective action. In the analysis made in \S
2
we have considered the spatial infinity as the only boundary.
Unless we  include a
source term, however, we have to add the contribution from the inner
boundary
into eq.(\ref{eq:CSac}). The result of substituting (\ref{eq:Gsol})
into
(\ref{eq:CSac}) turns out to be constant even if we do not impose
eq.(\ref{eq:eom}), which should be derived from the variational
principle.
To look for the origin of this inconsistency, we consider the variation
of the action:
$$
\delta I=2\int_{M}{\rm Tr}[\delta A\wedge(dA+A\wedge A)].
$$
{}From (\ref{eq:Gsol}) we see that $F=dA+A\wedge A$ is proportional
to $dt\wedge d\phi$ with nonvanishing terms being the coefficients of
$J_{0},J_{2},P_{0}$ and of $P_{2}$.  To obtain the equations of motion
(\ref{eq:eom}), therefore, we need nontrivial values of
$e^{0}_{r},e^{2}_{r},
\omega^{12}_{r}$ or of $\omega^{01}_{r}$, which vanish in our gauge.
This result is closely related to the fact that, in pure Chern-Simons
gauge
theories on a cylinder, the moduli space of flat connections does not
have
any nontrivial symplectic structure.
To obtain a nontrivial dynamics, we have to take account of  the gauge
degrees of freedom, whose values on the boundary become
dynamical degrees of freedom described by Chiral Wess-Zumino-Witten
action \cite{elitz}.

In our formulation we should first introduce the \lq\lq small" gauge
degrees
of freedom. The relevant connection is:
$$
g_{s}Ag_{s}^{-1}-dg_{s}g_{s}^{-1},\quad
g_{s}=\exp(P_{a}\zeta^{a}+J_{0}\eta^{0}+J_{2}\eta^{2}),
$$
where $A$ is given by (\ref{eq:Gsol}), and
$\zeta^{a},\eta^{0},\eta^{2}\sim
O(\frac{1}{r^{3}})$ are the small gauge degrees of freedom
which are subject to some boundary condition imposed
on an appropriately chosen inner boundary. While it is desirable to
obtain
the action which reproduce the equations of motion (\ref{eq:eom}) and
which is invariant under the transformation (\ref{eq:virasoro}), it is
probable
that the boundary condition on the inner boundary violates the symmetry
under
(\ref{eq:virasoro}), leaving only a subgroup as the symmetry of the
system.
\footnote{Recently WZW theory of 3DBH's has been
formulated in the viewpoint of the black hole
statistical mechanics\cite{carlip}. In that case, due to the boundary
condition
imposed on the outer horizon, the gauge symmetry group on the horizon
is reduced to the group of rigid rotations.}

\vskip2.5cm

\noindent Acknowledgments

I would like to thank Prof. K. Kikkawa, Prof. H. Itoyama and H.
Kunitomo for helpful discussions and
careful readings of the manuscript.
I am also grateful to Prof. T. Kubota for letting me aware of some
useful literature. The author is supported by the Japan Society for the
Promotion of Science.


\newpage
{\bf Figure Captions}
\begin{description}
\item[Fig.1] \  Behavior of the conventional radial coordinate
$\tilde{r}$
in the complex $\zeta$-plane (in the case $M^{+},M^{-}>0$).
\item[Fig.2] \  (a)Tangent space of ${\cal Q}$ at the
\lq\lq triple stabilizer point". (b) Topology of the space ${\cal Q}$
of
centrally extended quadratic differentials. The solid line,
dotted lines, and  dots  denote the spaces of
$T$ stabilized by type O)-, type I)- and type II)-stabilizers
respectively.

\item[Fig.3] \  Topology of the moduli space ${\cal C}$ of flat
$\tilde{SL}(2,{\bf R})$ connections on a cylinder.  The part drawn by
bold lines is homeomorphic to ${\cal Q}$.

\end{description}

\end{document}